\documentclass[aps,prc,twocolumn,showpacs,floatfix]{revtex4-1}
\usepackage{amssymb}
\usepackage{amsmath,bm}
\usepackage[normalem]{ulem}
\usepackage[dvips]{color}
\usepackage{booktabs}
\usepackage{graphicx}
\usepackage{float}
\usepackage{multirow}
\usepackage{lineno}
\usepackage[normalem]{ulem}
\usepackage[usenames, dvipsnames]{xcolor}
\usepackage{tensor}
\usepackage[utf8]{inputenc}
\usepackage{subfigure}
\begin{document}

\newcommand{\hl}{$\rm ^4_{\Lambda}H$~}
\newcommand{\hld}{$\rm ^4_{\Lambda}H$}
\newcommand{\hel}{$\rm ^4_{\Lambda}He$~}
\newcommand{\held}{$\rm ^4_{\Lambda}He$}
\newcommand{\he}{$\rm ^3He$~}
\newcommand{\la}{$\rm \Lambda$~}

\newcommand{\com}[1]{{\sf\color[rgb]{0,0,1}{#1}}}
\newcommand{\sNN}[1]{$\sqrt{s_\mathrm{NN}}$~}



\title[mode = title]{Measurement of \hl and \hel binding energy in Au+Au collisions at \sNN~=~3 GeV}

\author{
M.~S.~Abdallah$^{4}$,
B.~E.~Aboona$^{53}$,
J.~Adam$^{6}$,
L.~Adamczyk$^{2}$,
J.~R.~Adams$^{38}$,
J.~K.~Adkins$^{30}$,
I.~Aggarwal$^{40}$,
M.~M.~Aggarwal$^{40}$,
Z.~Ahammed$^{59}$,
D.~M.~Anderson$^{53}$
% ...
}

\address{\rm{(STAR Collaboration)}}

% ...
\address{$^{2}$AGH University of Science and Technology, FPACS, Cracow 30-059, Poland}
% ...
\address{$^{4}$American University of Cairo, New Cairo 11835, New Cairo, Egypt}
% ...
\address{$^{6}$Brookhaven National Laboratory, Upton, New York 11973}
% ...
\address{$^{30}$University of Kentucky, Lexington, Kentucky 40506-0055}
% ...
\address{$^{38}$Ohio State University, Columbus, Ohio 43210}
% ...
\address{$^{40}$Panjab University, Chandigarh 160014, India}
% ...
\address{$^{53}$Texas A\&M University, College Station, Texas 77843}
% ...
\address{$^{59}$Variable Energy Cyclotron Centre, Kolkata 700064, India}
% ...

\date{\today}

\begin{abstract}
    Measurements of mass and $\Lambda$ binding energy of \hl and \hel in Au+Au collisions at $\sqrt{s_{_{\rm NN}}}=3$ GeV are presented, with an aim to address the charge symmetry breaking (CSB) problem in hypernuclei systems with atomic number A = 4. The $\Lambda$ binding energies are measured to be $\rm 2.22\pm0.06(stat.) \pm0.14(syst.)$~MeV and $\rm 2.38\pm0.13(stat.) \pm0.12(syst.)$~MeV for \hl and \held, respectively. The measured $\Lambda$ binding-energy difference is $\rm 0.16\pm0.14(stat.)\pm0.10(syst.)$~MeV for ground states. Combined with the $\gamma$-ray transition energies, the binding-energy difference for excited states is $\rm -0.16\pm0.14(stat.)\pm0.10(syst.)$~MeV, which is negative and comparable to the value of the ground states within uncertainties. These new measurements on the $\Lambda$ binding-energy difference in A = 4 hypernuclei systems are consistent with the theoretical calculations that result in $\rm \Delta B_{\Lambda}^4(1_{exc}^{+})\approx -\Delta B_{\Lambda}^4(0_{g.s.}^{+})<0$ and present a new method for the study of CSB effect using relativistic heavy-ion collisions.
    \end{abstract}
    
    \maketitle
    
    \section{Introduction}
    
    Nuclei containing strange quarks, called hypernuclei, are ideal hyperon-baryon bound systems for studying the hyperon-nucleon (YN) interactions and have therefore been the subject of intense study~\cite{STAR:2017gxa,ALICE:2019vlx,Chen:2018tnh,STAR:2010gyg}. The $\rm \Lambda$ binding energy $\rm B_{\Lambda}$ (also called the \la separation energy) of a hypernucleus is defined as the difference between the mass of the hypernucleus, and the sum of the masses of the nucleon core and the $\Lambda$:
    \begin{align}
            B_{\Lambda} = (M_{\Lambda} + M_{\rm core} - M_{\rm hypernucleus})c^2.
        \label{eq1}
    \end{align}
    The determination of \la binding energies can aid in the understanding of YN interactions and the equation of state (EOS) of hypernuclear matter with a potential connection to neutron star studies~\cite{Lonardoni:2014bwa,Fortin:2017cvt}. And it has been the subject of theoretical calculations and experimental measurements~\cite{Nogga:2001ef,Tamura:2019yxd,Liu:2019mlm,Abdurakhimov:1989jk}. Recent results from the STAR Collaboration~\cite{STAR:2019wjm} have shown the \la binding energy of the hypertriton to be larger than zero, challenging previous results~\cite{Juric:1973zq}. Precision measurements of \la binding energies of heavier hypernuclei than the hypertriton are expected to improve our understanding of the YN interactions between $\Lambda$ and heavier nuclei.
    
    \begin{figure*}[!htbp]
        \centering
        \includegraphics[width=0.8\textwidth]{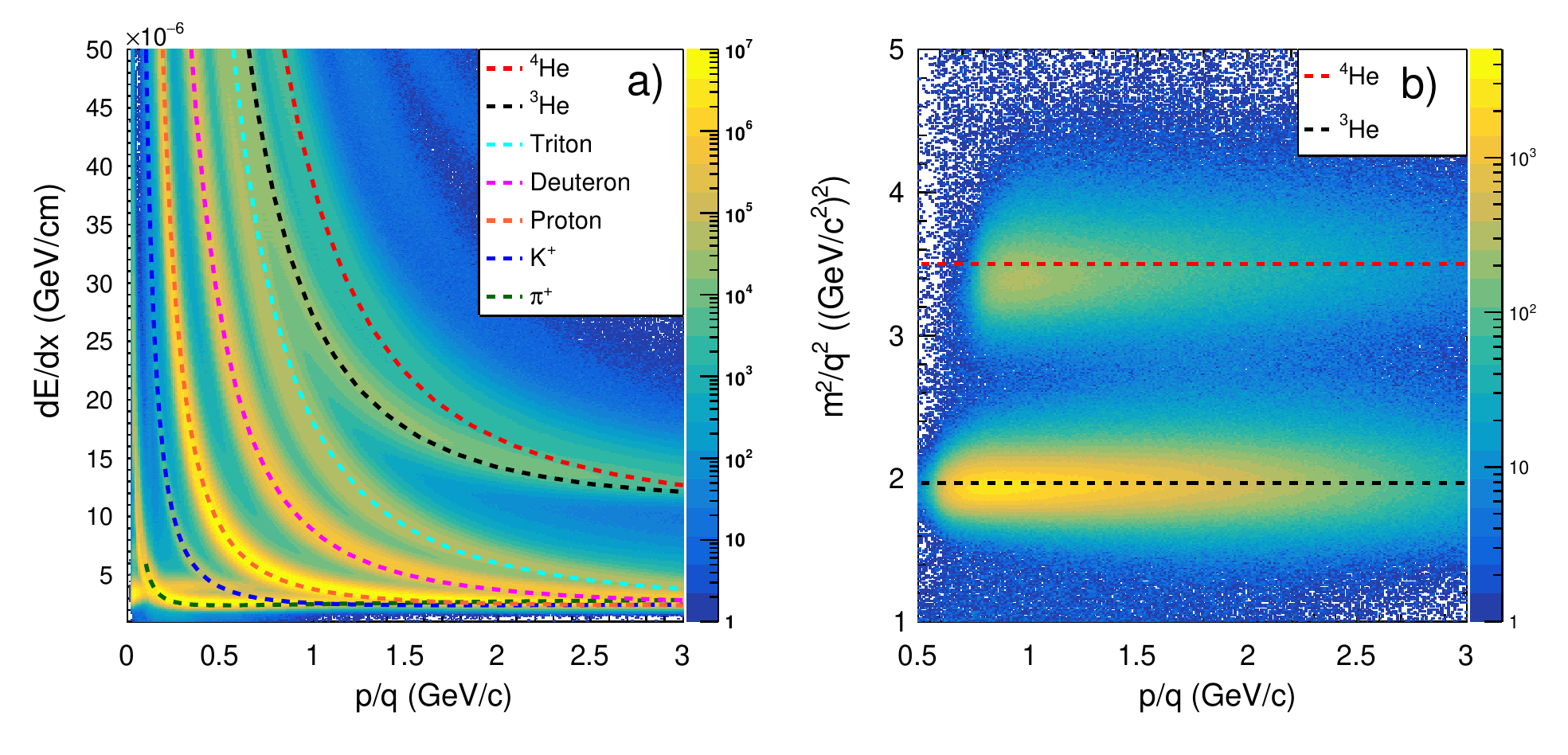}
        \caption{(a): The mean energy loss in the TPC versus rigidity, $p/q$, where $p$ is the momentum of the particle and $q$ is its electric charge in units of the electron charge. The dashed curves represent the expected values calculated by the Bichsel function~\cite{Bichsel:2006cs} for each particle species. (b): The square of the ratio of mass and charge, $m^2/q^2$, versus rigidity in the TOF detector. The dashed curves represent the expected values for $\rm ^{3}He$ and $\rm ^{4}He$.}
        \label{fig1}       
    \end{figure*}
    
    The charge symmetry of the strong interaction predicts that the $\Lambda p$ and the $\Lambda n$ interaction should be identical, because $\Lambda$ is charge neutral. The binding-energy difference between a pair of mirror nuclei, whose numbers of protons and neutrons are exchanged, originates from the difference of the Coulomb interactions and the mass difference of the up and down quarks~\cite{Machleidt:2000vh}. Furthermore, the $\Lambda$ binding energy of mirror hypernuclei such as \hl (triton + $\Lambda$) and \hel ($^3$He + $\Lambda$) should be equal according to charge symmetry. However, the measured difference in binding energy between the triton and $\rm ^3He$ demonstrates the breaking of charge symmetry. With the removal of the contributions from Coulomb interactions, the value of the binding energy difference between the triton and $\rm ^3He$ is 67 $\pm$ 9~keV~\cite{Machleidt:2000vh}. On the other hand, measurements in nuclear emulsion experiments reported a $\Lambda$ binding-energy difference $\Delta B_{\Lambda}^4(0_{g.s.}^{+}) = 350 \pm 50~$~keV~\cite{Juric:1973zq} between \hl and \hel in their ground states, which is larger than the binding-energy difference in nuclei, representing a puzzle since reported~\cite{Juric:1973zq}.
    
    In 2015, the J-PARC E13 $\gamma$-ray spectroscopy experiment measured the $\gamma$-ray transition energy for the $1^{+}$ first excited state of \hel to be $\rm 1406~\pm~2(stat.)~\pm~2(syst.)~keV$~\cite{Yamamoto:2015avw}. The E13 Collaboration then combined the $\Lambda$ binding energies of ground states from emulsion experiments in the 1970s~\cite{Juric:1973zq}, the $\gamma$-ray transition energy for \hl measured in 1976~\cite{Bedjidian:1976zh}, and their new $\gamma$-ray transition energy measurement for \hel to determine the difference in excited states as $\rm \Delta B_{\Lambda}^4(1_{exc}^{+}) = 30 \pm 50$~keV~\cite{Yamamoto:2015avw}. This is roughly a factor of ten smaller than that in the ground states~\cite{Juric:1973zq}. It was also suggested that the CSB effect may have a significant spin dependence which is larger in ground states than in excited states~\cite{Yamamoto:2015avw}. In 2016, the A1 Collaboration at the Mainz Microtron used spectrometers to make a new measurement of the ground state $\Lambda$ binding energy of \hld~\cite{Esser:2015trs,Schulz:2016kdc}. Combining their new measurement with the previous $\Lambda$ binding energy of \held~\cite{Juric:1973zq} and the measurements of the $\gamma$-ray transition energies for \hld~\cite{Bedjidian:1976zh} and \held~\cite{Yamamoto:2015avw}, the binding-energy differences were updated to be $\rm \Delta B_{\Lambda}^4(0_{g.s.}^{+}) = 233 \pm 92$~keV and $\rm \Delta B_{\Lambda}^4(1_{exc}^{+}) = -83 \pm 94$~keV~\cite{Esser:2015trs,Schulz:2016kdc}. 
    
    Many theoretical model calculations have failed to reproduce the experimental results, with most of them underestimating the CSB effect in both the ground and excited states~\cite{Nogga:2001ef,Haidenbauer:2007ra,Nogga:2013pwa,Coon:1998jd}. It has been proposed that $\Lambda-\Sigma$ mixing can account for the large CSB~\cite{Gal:2015bfa}. In 2016, the $ab$ $initio$ calculation using chiral effective field theory hyperon-nucleon potentials plus a CSB $\Lambda-\Sigma^0$ mixing vertex of A = 4 hypernuclei achieved a large CSB in both ground and excited states, and also concluded that $\rm \Delta B_{\Lambda}^4(1_{exc}^{+})\approx -\Delta B_{\Lambda}^4(0_{g.s.}^{+})<0$~\cite{Gazda:2015qyt}. Independent experiments are needed to test these calculations. More accurate values of the $\Lambda$ binding-energy splitting in ground and excited states are needed to constrain the $\Lambda$n interaction~\cite{Haidenbauer:2021wld}.

    To study the QCD matter in the high-baryon-density region, the STAR detector acquired data for collisions at the lowest available energy of the BES-II program. In 2018, STAR collected over $3\times10^8$ events at a center-of-mass energy of \sNN~=~3~GeV. The UrQMD-hydro hybrid model predicts that the production yields of hypernuclei is at a maximum around \sNN~=~5~GeV with high baryon chemical potential~\cite{Steinheimer:2012tb}. Therefore, \sNN~=~3~GeV collisions collected with the STAR experiment provide an opportunity to study the $\Lambda$ binding energies of \hl and \hel in the same experiment to address the CSB problem.  
    
    \section{Analysis Details}\label{sec:II}
    \subsection{The STAR Detector}\label{sec:IIA}
    This work is based on a high-statistics data set of Au+Au collisions at $\sqrt{s_{\rm NN}}=3~$GeV taken in fixed-target mode using the STAR detector in 2018. A 0.25~mm thick stationary gold target was mounted inside the beam pipe 2~m to the center of the Time Projection Chamber (TPC)~\cite{Anderson:2003ur}. In the collider mode, the lowest $\sqrt{s_{\rm NN}}$ for Au+Au collisions that RHIC can run with usable luminosity is 7.7 GeV, whereas in the fixed-target mode this low energy limit can be extended to 3 GeV. A gold beam incident from the same side as the gold target at laboratory kinetic energy 3.85$A$~GeV produces collisions at $\sqrt{s_{\rm NN}}=3~$GeV in the center-of-mass frame. The collision vertices are selected to be within 2~cm of the gold target's position in the longitudinal (beam) direction and also within 2~cm of the average position of collision vertices in the transverse plane. With these selections, 317 million events with minimum bias trigger~\cite{STAR:2021orx} are analyzed in this paper.
    
    The particle identification (PID) is achieved with the TPC and the Time-of-Flight (TOF) detector~\cite{Llope:2012zz}. The TPC data allow the reconstruction of the paths of emitted particles and provides particle identification via the measurement of energy loss, $dE/dx$. Figure~\ref{fig1}(a) presents the distribution of tracks versus $dE/dx$ and magnetic rigidity, $p/q$, using the TPC. A 0.5~T magnetic field is applied along the TPC's cylindrical axis causing the charged tracks to follow helical paths, the curvatures of which reveal the track rigidity. The dotted curves are calculations of the Bichsel function~\cite{Bichsel:2006cs} for the indicated particle species. The PID for $\pi^-$, proton, $^3$He, and $^4$He are firstly achieved by selecting the measured $dE/dx$ within 3 standard deviations of their expected values by Bichsel function. These tracks are also required to have more than 15 space points in the TPC.
    
    As seen in Fig.~\ref{fig1}(a), the particle species are not completely separated by the TPC. The TOF detector measures a particle's time of travel from the collision vertex to the TOF location, and offers species separation to higher momentum than $dE/dx$ alone. As evident from Fig.~\ref{fig1}(b), $\rm ^{3}He$ and $\rm ^{4}He$ are separated clearly. By selecting the $\rm ^{3}He$ and $\rm ^{4}He$ tracks within the ranges from 1.4 to 2.5 (GeV/$c^2$)$^2$ and from 2.5 to 4.5 (GeV/$c^2$)$^2$ of their $m^2/q^2$ respectively, their purities can both reach to 95\%. This information is only used in the identification of $\rm ^{3}He$ and $\rm ^{4}He$ when the relevant TOF signals are matched to TPC tracks. Otherwise only the TPC information is used.
    
    \begin{figure*}[!htbp]
        \centering
        \includegraphics[width=0.9\textwidth]{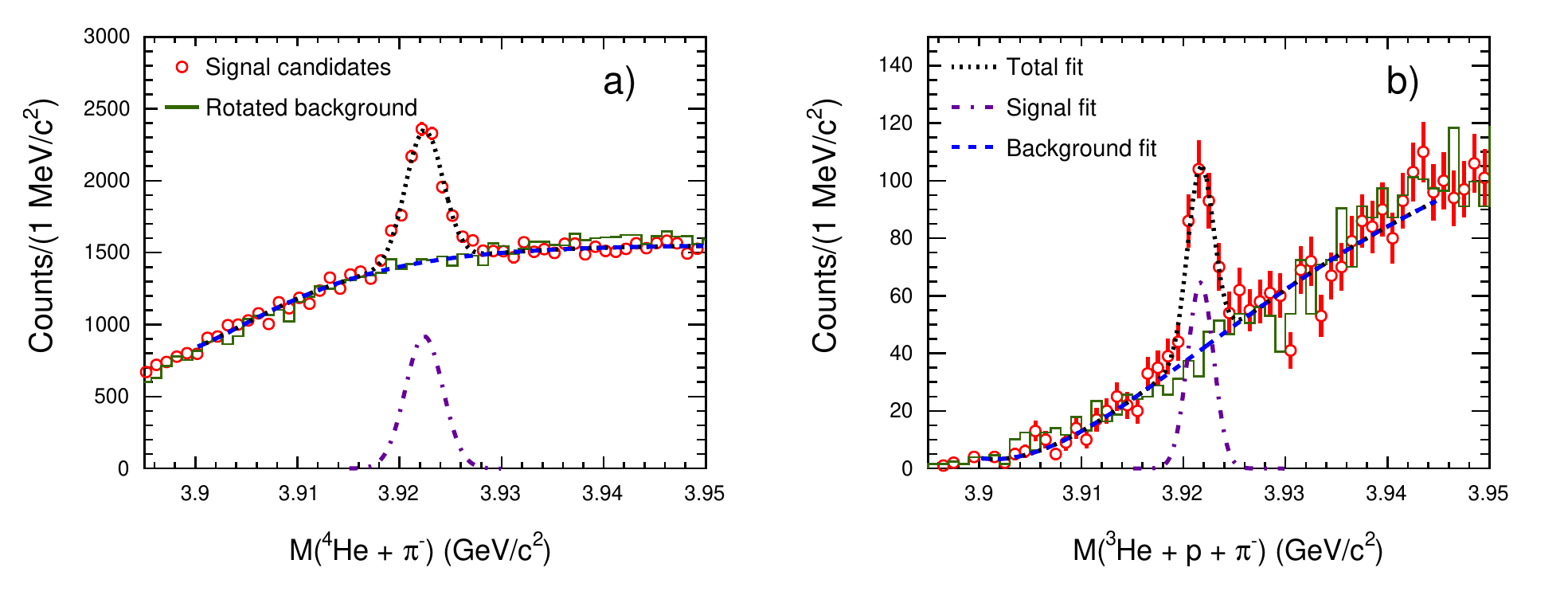}
        \caption{Invariant-mass distributions for \hl (a) and \hel (b) reconstructed with KFParticle and TMVA-BDT. The green histograms represent the rotated backgrounds. The blue dashed curves represent the background fits and are obtained by fitting the invariant-mass distributions outside of the signal regions with double-exponential functions. The black dashed curves are obtained by fitting these distributions across the full range of invariant mass with the background fit result and a Gaussian function. The violet dashed curves represent the signal Gaussian functions.}
        \label{fig2}       
    \end{figure*}

    \subsection{Signal reconstruction}\label{sec:IIB}
    
    In this analysis, the \hl is reconstructed via its two-body decay channel, $\rm ^4_{\Lambda}H\to~^4He+\pi^-$, and \hel is reconstructed via its three-body decay channel, $\rm ^4_{\Lambda}He\to~^3He+p+\pi^-$. The discussion on \hl three-body decay channel can be found in Section~\ref{sec:III}. The daughter particles are identified according to the methods described in Section~\ref{sec:IIA}. The KFParticle package~\cite{Kisel:2020lpa,Zyzak2016} is used to reconstruct the invariant-mass distributions of \hl and \held. KFParticle package is an algorithm based on the Kalman filter to reconstruct short-lived particles in heavy-ion collisions~\cite{STAR:2021orx}. In KFParticle, a particle is described by a state vector constructed by its coordinate and momentum information from the detector and a covariance matrix associated with the state vector. Various topological variables, including the distance of closest approach (DCA) between a particle and the primary vertex (PV) and DCA between the decay daughters, are used to suppress the background. In KFParticle, the DCA can also be represented by the covariance between two points, $\chi^2$, calculated by the covariance matrix of the track. Smaller value of $\chi^2$ corresponds to a closer distance. With the decay daughters identified, the invariant-mass distributions of hypernuclei can be determined.
    
    To optimize the signal, the TMVA-BDT~\cite{Hocker:2007ht} package is used. The Boosted Decision Trees (BDT) algorithm can distinguish signal from background according to topological variables. In this analysis, six topological variables are used as training features for \hld: the decay length of \hld, the decay length over its error calculated by the covariance matrix, the $\chi^2$ of the DCA between \hl and the PV, the $\chi^2$ of the DCA between decay daughters, the $\chi^2$ of the DCA between $\pi^-$ and the PV, and the DCA between $\pi^-$ and the PV.  For \held, five topological variables are used: the \hel decay length, the $\chi^2$ of the DCA between \hel and the PV, the $\chi^2$ of the DCA between the decay daughters, the $\chi^2$ of the DCA between the proton and the PV, and the $\chi^2$ of the DCA between $\pi^-$ and the PV. The BDT algorithm is trained to calculate a response value for each candidate to distinguish signal and background. The reconstructed particles from simulated events are used as the training sample for signals. Here the \hl and \hel particles are simulated using the GEANT software~\cite{GEANT4:2002zbu} with STAR detector geometry and materials. The output detector responses are embedded into real data samples, then reconstructed just like real data. The samples for background are obtained from the real experimental data by rotating the $\rm ^4He$ or $\rm ^3He$ track by 180 degrees around the longitudinal axis before applying the reconstruction method. Panels (a) and (b) in Fig.~\ref{fig2} show the invariant-mass distributions alongside fittings to the signal and background regions of \hl and \hel reconstructed with KFParticle and TMVA-BDT optimization. Here, we correct for the effects of energy loss and magnetic field measurement inaccuracy on the measured momenta of decay daughters. These corrections will be discussed in Section~\ref{sec:IIC}.  We define significance $S/\sqrt{S+B}$, where $S$ and $B$ are the counts of signal and background, respectively, in the invariant mass region. The significances for \hl and \hel are about 36 and 10, respectively.
    
    As a cross check of the reconstruction algorithm for \hl and \held, a ``helix swimming'' method~\cite{STAR:2002fhx,STAR:2019wjm} to find the closest approach among daughters is also implemented. By tuning topological variable cuts and the optimization of TMVA-BDT, the \hl and \hel mass results from helix swimming are consistent with those from KFParticle with mass difference at the level of 10~keV. 
    
    \subsection{Corrections}\label{sec:IIC}
\begin{figure}[!htbp]
        \centering
        \includegraphics[width=0.5\textwidth]{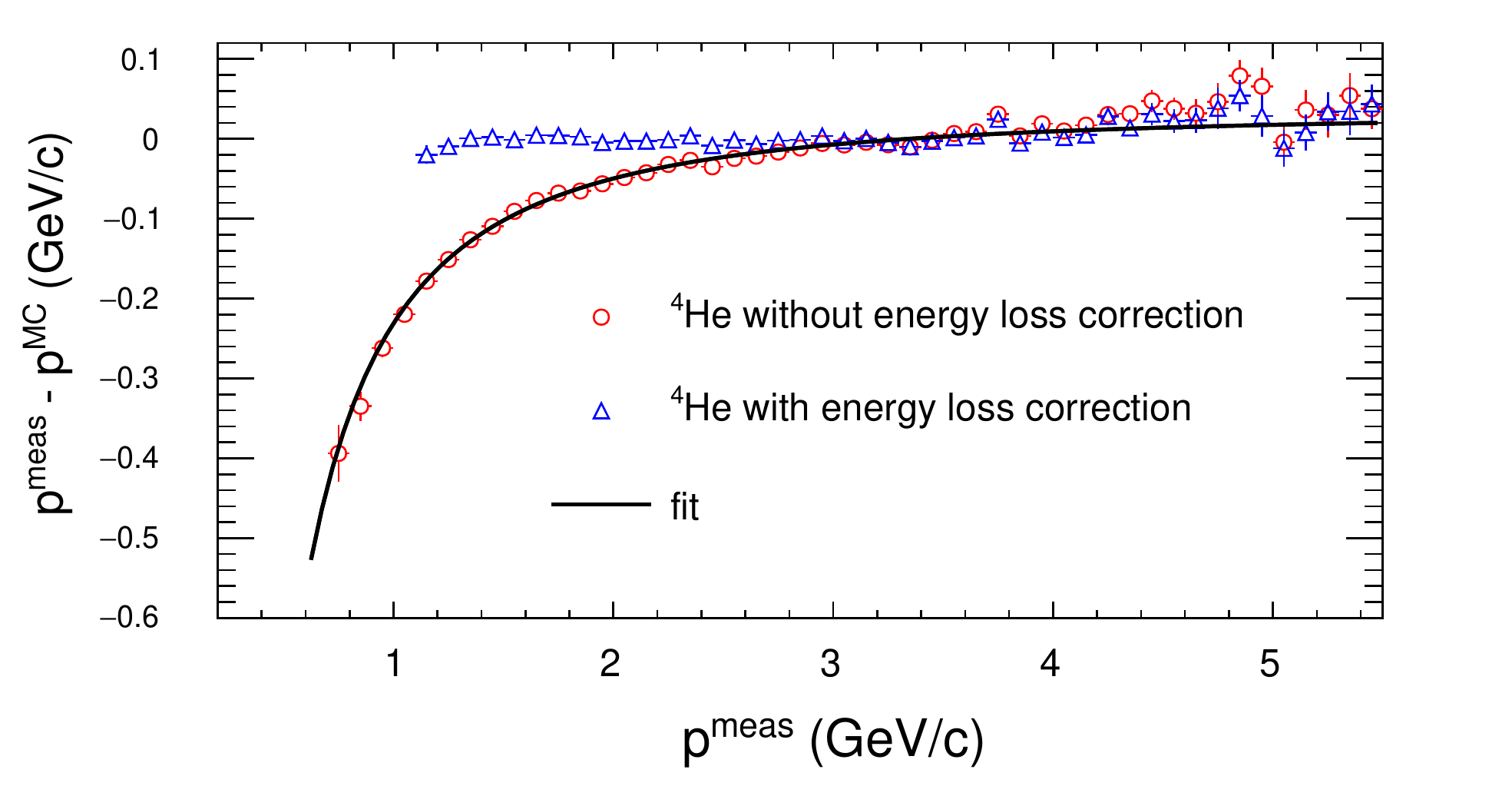}
        \caption{The average difference between the measured momentum and the MC input momentum as a function of the measured momentum for $\rm ^4He$. The red circles represent the energy loss without any corrections and the black curve is the fit for them. The blue triangles are the energy loss with energy-loss correction applied on the measured momentum. 
        }
        \label{fig_eloss}       
    \end{figure}

    Particles emitted from the collisions lose energy in a momentum-dependent manner when passing through materials before entering the tracking region of the TPC. This effect necessitates an appropriate energy-loss correction on track momenta. During the track-reconstruction process, the energy-loss effect is considered assuming that all particles are pions. So it is necessary here to apply additional energy-loss corrections for $\rm ^4He$, $\rm ^3He$ and proton. Similar to the method performed in Ref.~\cite{STAR:2008med}, STAR simulation and embedding for \hl and \hel samples are used to study these corrections. By comparing the difference between the measured momentum magnitude $p^{\rm meas}$ and the Monte Carlo (MC) input momentum magnitude $p^{\rm MC}$, the momentum-loss effect as a function of the $p^{\rm meas}$ can be determined. The red circles in Fig.~\ref{fig_eloss} represent the average momentum-loss effect versus $p^{\rm meas}$ of $\rm ^4He$ as an example. It is clear that the momentum-loss effect for $\rm ^4He$ is significant in the low-momentum region.
    
    The momentum-loss effect versus the measured momentum can be fitted with the correction function:
    \begin{align}
        p^{\rm meas}-p^{\rm MC}=\delta_0+\delta\left(1+\frac{m^2}{(p^{\rm meas})^2}\right)^{\alpha},
        \label{eloss_fit}
    \end{align}
    where $m$ is the mass of the particle and $\delta_0$, $\delta$, and $\alpha$ are fitting parameters. The fit results shown in Table~\ref{tab0} are then used to correct the momenta of $\rm ^4He$, $\rm ^3He$, and proton before performing the \hl and \hel reconstruction.
    
    \begin{table}[!htbp]
        \centering
        \caption{The values and the corresponding statistical uncertainties of fitting parameters used for the energy loss corrections for $^4$He, $^3$He, and proton.}
        \label{tab0}       
        \setlength{\tabcolsep}{2.5mm}{
        \begin{tabular}{cccc}
        \hline
        \hline
        Particle & $\delta_0$       & $\delta$      & $\alpha$ \\\hline
        $^4$He   & 0.072$\pm$0.007 & -0.039$\pm$0.005 &0.757$\pm$0.040 \\
        $^3$He & 0.036$\pm$0.003 & -0.020$\pm$0.002 &0.882$\pm$0.039 \\
        proton    & 0.024$\pm$0.002 & -0.021$\pm$0.002 &0.396$\pm$0.027 \\
        \hline
        \hline
        \end{tabular}}
    \end{table}
    
    Another correction is applied when we verify that the correct $\Lambda$ mass is reconstructed. All track momenta are scaled by the factor 0.998 to make the measured mass of $\Lambda$ match the PDG value~\cite{ParticleDataGroup:2020ssz}. This discrepancy could be caused by differences between the true and nominal current which controls the magnetic field strength in STAR detector. With this correction, the invariant-mass distribution of reconstructed $\Lambda$, which is discussed in Section~\ref{sec:IID}, is peaked at the appropriate PDG value~\cite{ParticleDataGroup:2020ssz}. 
    
    \subsection{Systematic uncertainties}\label{sec:IID}
    Since the uncertainties on the masses of $\Lambda$, triton, and $^3$He used in the calculations for $\Lambda$ binding energies are quite small~\cite{ParticleDataGroup:2020ssz,CPC41}, the systematic uncertainties for the $\Lambda$ binding energies are the same as them for the measured masses of the hypernuclei in this analysis. These systematic uncertainties mainly come from the aforementioned corrections. For the energy loss corrections, the correction parameters with their statistical uncertainties $\sigma$ are obtained from the fits with Eq.~(\ref{eloss_fit}). The parameters are varied from $+1\sigma$ to $-1\sigma$ to investigate their influences on the measurements. The average difference of the measurements with these variations are taken as the systematic uncertainties.
    
    The systematic uncertainty of the momentum scaling factor 0.998 is evaluated by measuring the $\Lambda$ hyperon mass via its two body decay channel $\Lambda \to  p~+~\pi^-$ in the same data set. With the energy-loss correction for the proton and the momentum scaling factor being applied, the extracted $\Lambda$ mass is still a function of its momentum, but remains within 0.10~MeV/$c^2$ of the PDG value 1115.683~$\pm$~0.006~MeV/$c^2$~\cite{ParticleDataGroup:2020ssz}. Thus, the 0.10~MeV/$c^2$ difference is propagated to the systematic uncertainties for \hl and \hel by scaling it with the ratio of the difference between the hypernuclei masses with and without the 0.998 scaling factor to the difference between the $\Lambda$ masses with and without the 0.998 factor. The resulting systematic uncertainties for \hl and \hel masses are both calculated to be 0.11~MeV/$c^2$.
    
    Variations of the measured mass by the change of BDT response cuts are also considered as a source of systematic uncertainty. The BDT response cut was varied in a large range and the final mass result is the average value of several fitting results of the invariant mass distributions with different cuts. The half of the maximum change in the mass is regarded as the systematic uncertainty. We also checked the fit of the signal after the combinatorial background was subtracted via the rotational-background method and found that the changes in the results are negligible. Table~\ref{tab1} summarizes the systematic uncertainties from various sources for \hl and \held. 
    
    \begin{table}[!htbp]
        \centering
        \caption{Sources of systematic uncertainties for the masses and $\Lambda$ binding energies of \hl and \hel in MeV/$c^2$.}
        \label{tab1}       
        \setlength{\tabcolsep}{6mm}{
        \begin{tabular}{ccc}
        \hline
        \hline
        Uncertainty source & \hl  & \hel   \\\hline
        Momentum scaling factor & 0.11 & 0.11 \\
        Energy loss correction & 0.08 & 0.05 \\
        BDT response cut    &0.03     & 0.01 \\
        Total                  & 0.14       & 0.12       \\
        \hline
        \hline
        \end{tabular}}
    \end{table}
    
    When measuring the $\Lambda$ binding-energy difference between \hl and \held, the systematic uncertainties from the momentum scaling factor will largely be canceled out, but the cancellation will not be complete due to their different decay phase spaces. We applied the 0.998 factor in the simulation data and found that it brings a 0.02~MeV change to the $\Lambda$ binding-energy difference. Thus this 0.02~MeV is considered as a systematic uncertainty for the $\Lambda$ binding-energy difference. The systematic uncertainties from other sources are added in quadrature to obtain the total systematic uncertainties of the $\Lambda$ binding-energy difference, summarized in Table~\ref{tab2}.
    
    \begin{table}[!htbp]
        \centering
        \caption{Systematic uncertainties for the difference of $\Lambda$ binding energies between \hl and \hel in the ground state in MeV.}
        \label{tab2}       
        \setlength{\tabcolsep}{4mm}{
        \begin{tabular}{ccc}
        \hline
        \hline
        Uncertainty source   &Uncertainty \\
        \hline
        Momentum scaling factor         &0.02  \\
        Energy loss correction          &0.09  \\
        BDT response cut                 &0.03  \\
        Total                           &0.10  \\
        \hline
        \hline
        \end{tabular}}
    \end{table}
    
    \section{Results and discussions}\label{sec:III}
\begin{figure}[!htbp]
        \centering
        \includegraphics[width=0.5\textwidth]{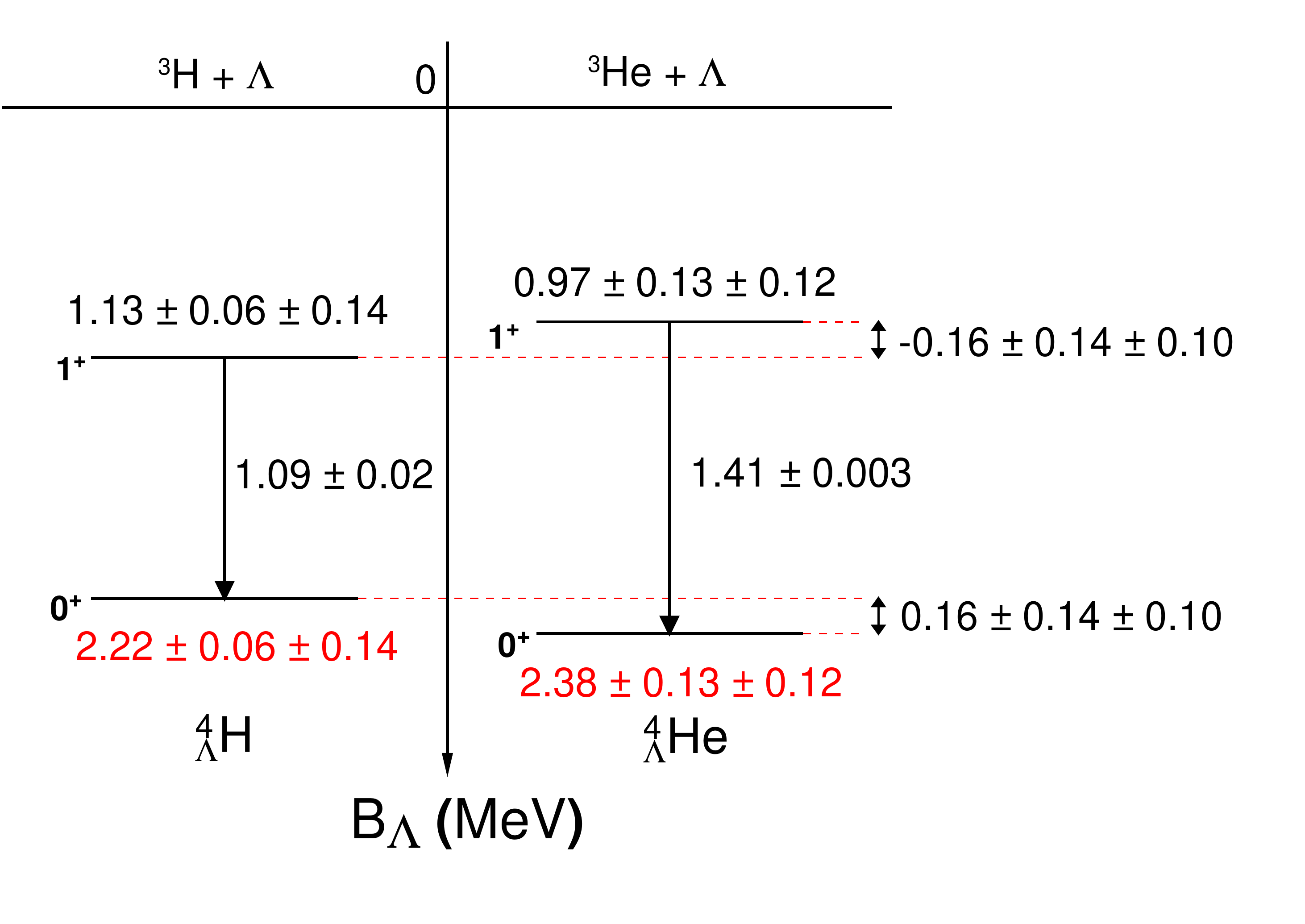}
        \caption{Energy level schemes of \hl and \hel in terms of $\Lambda$ binding energies. The ground-state binding energies are from this analysis. The values for excited states are obtained from the $\gamma$-ray transition energies measured in Refs.~\cite{Bedjidian:1976zh,Yamamoto:2015avw}. 
        }
        \label{fig3}       
    \end{figure}

    The signal and the background in the invariant-mass distributions of \hl and \hel are fitted by a Gaussian distribution and a double-exponential function, respectively:
    \begin{align}
        f(x) =& \frac{A}{\sqrt{2\pi}\sigma}\exp\left(-\frac{(x-\mu)^2}{2\sigma^2}\right)+p_0\exp\left(-\frac{x-p_1}{p_2}\right) \nonumber \\
        &+p_3\exp\left(-\frac{x-p_1}{p_4}\right) + p_5 .
    \end{align}
    The fitting result of $\mu$ is the mass of the interested hypernucleus. The fitting results are shown as the black dashed curves in Fig.~\ref{fig2}. Using the methods which has been described in Section~\ref{sec:II}, we have measured $m$$\rm (^4_{\Lambda}H)~=~3922.38 \pm 0.06(stat.) \pm 0.14(syst.)~MeV/$$c^2$, and $m$$\rm (^4_{\Lambda}He)~=~3921.69 \pm 0.13(stat.) \pm 0.12(syst.)~MeV/$$c^2$. We can extract the $\Lambda$ binding energies of \hl and \hel according to Eq.~\ref{eq1}. The mass of $\Lambda$ ($m$$(\Lambda) = 1115.68$~MeV/$c^2$) is taken from the PDG~\cite{ParticleDataGroup:2020ssz}, and the masses of triton ($m$$(t) = 2808.92$~MeV/$c^2$) and $\rm ^3He$ ($m$$(\rm ^3He) = 2808.39$~MeV/$c^2$) are from CODATA~\cite{CPC41}. With the mass measurements in this analysis, the $\Lambda$ binding energies of \hl and \hel are $B$$\rm _{\Lambda}(^4_{\Lambda}H)~=~2.22 \pm 0.06(stat.) \pm 0.14(syst.)~MeV$ and $B$$\rm _{\Lambda}(^4_{\Lambda}He)~=~2.38 \pm 0.13(stat.) \pm 0.12(syst.)~MeV$. These results are illustrated in Fig.~\ref{fig3}.

    
    The $\Lambda$ binding energies of \hl and \hel in this analysis correspond to the ground states, reconstructed via their weak-decay channels. The $\Lambda$ binding energies in excited states can be obtained according to the $\gamma$-ray transition energies of the excited \hl and \held. Combined with the $\gamma$-ray transition energies obtained from previous measurements, $E$$\rm _{\gamma}(^4_{\Lambda}H)~=~1.09~\pm~0.02~MeV$~\cite{Bedjidian:1976zh} and $E$$\rm _{\gamma}(^4_{\Lambda}He)~=~1.406~\pm~0.003~MeV$~\cite{Yamamoto:2015avw}, the $\Lambda$ binding-energy differences between \hl and \hel are $\Delta B$$\rm _{\Lambda}^4(0_{g.s.}^{+})~=~\rm0.16 \pm 0.14(stat.) \pm 0.10(syst.)~MeV$ and $\Delta B$$\rm _{\Lambda}^4(1_{exc}^{+})~=~\rm-0.16 \pm 0.14(stat.) \pm 0.10(syst.)~MeV$.
    
    \begin{figure*}[!htbp]
        \centering
        \includegraphics[width=1.0\textwidth]{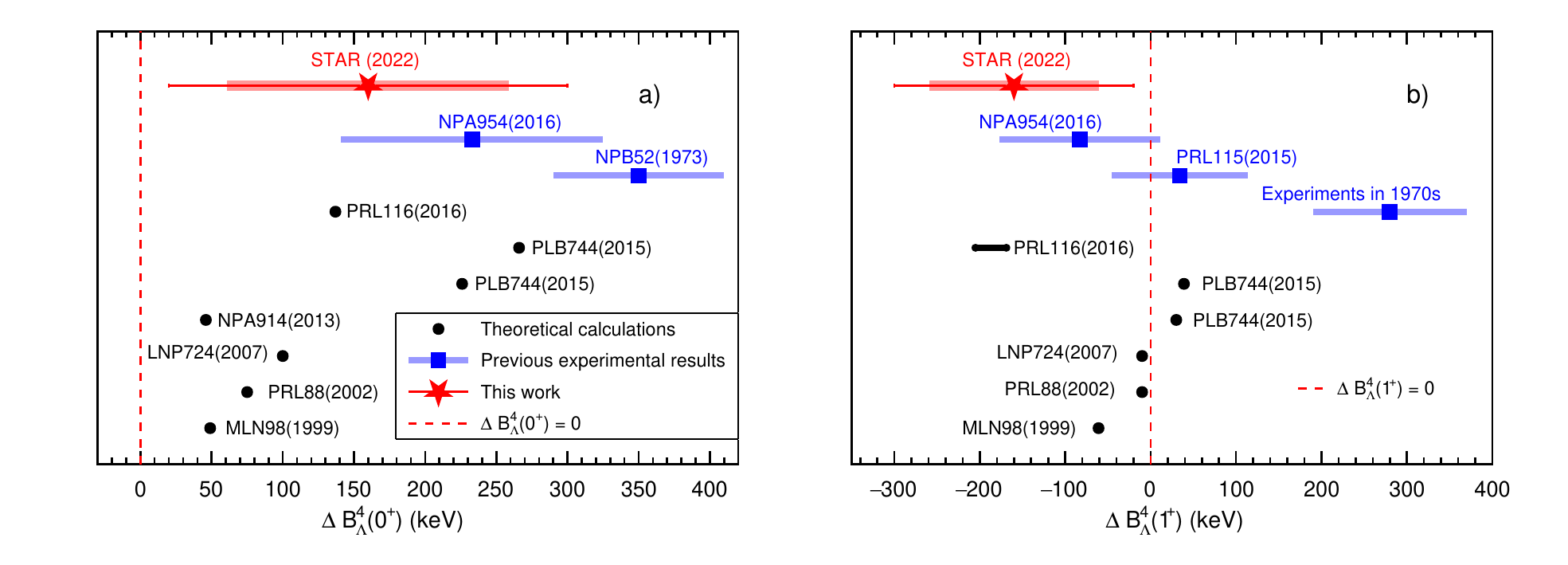}
        \caption{The $\Lambda$ binding-energy differences between \hl and \hel in ground states (a) and in excited states (b) compared with theoretical model calculations (black solid circles and a short black line) and previous measurements (blue solid squares). Solid error bars show statistical uncertainties and boxes show the systematic uncertainties. Red dashed vertical lines are drawn at $\rm \Delta B^4_{\Lambda}(0^+~or~1^+)=0$.
        }
        \label{fig4}       
    \end{figure*}
    
    Figure~\ref{fig4} presents a compilation of current measurements together with early measurements~\cite{Juric:1973zq,Yamamoto:2015avw,Esser:2015trs,Schulz:2016kdc,Bedjidian:1976zh,CERN-Lyon-Warsaw:1979ifx} and theoretical model calculations~\cite{Gal:2015bfa,Gazda:2015qyt,Nogga:2001ef,Haidenbauer:2007ra,Nogga:2013pwa,Coon:1998jd} for the $\Lambda$ binding-energy differences. The solid blue square markers in Fig.~\ref{fig4} show results from nuclear emulsion experiments in 1970s, in which a positive binding-energy difference in the excited states with a magnitude similar to the ground states was measured. This similarity arises because the $\gamma$-ray transition energy for \hel was measured to be $\rm \rm E_{\gamma}(^4_{\Lambda}He)~=~1.15~\pm~0.04~MeV$ at that time~\cite{,CERN-Lyon-Warsaw:1979ifx}, which is comparable to that of \hl\cite{Bedjidian:1976zh}. With a precise measurement of the $\gamma$-ray transition energy for \hel in 2015~\cite{Yamamoto:2015avw}, which shows a larger $\gamma$-ray transition energy for \hel than for \hld, the $\Lambda$ binding energy difference in excited states was calculated to be around zero, and it is much smaller than that in ground states. As discussed in the introduction and shown as solid black circle markers in Fig.~\ref{fig4} with black dots, most of the theoretical calculations predict small $\Lambda$ binding-energy differences in both ground states and excited states~\cite{Nogga:2001ef,Haidenbauer:2007ra,Nogga:2013pwa,Coon:1998jd}. Reference~\cite{Gazda:2015qyt} (denoted as PRL116(2016)) predicts large values of $\Lambda$ binding energy differences in both ground states and in excited states with opposite sign, i.e. $\rm \Delta B_{\Lambda}^4(1_{exc}^{+})\approx -\Delta B_{\Lambda}^4(0_{g.s.}^{+})$. Within current uncertainties, this prediction matches our measurements. This may indicate that the CSB effect is comparable and has the opposite sign in ground states and excited states in $A = 4$ hypernuclei which has not been shown in previous measurements. An accurate measurement of the $\gamma$-ray transition energy for excited \hl is important as it directly impacts the deduced $\Lambda$ binding energy for the excited state. Currently, our results are based on the $\gamma$-ray transition energy for \hl from the experiments in the 1970s which show a large difference from the recent measurements in the $\gamma$-ray transition energy for \held~\cite{CERN-Lyon-Warsaw:1979ifx,Yamamoto:2015avw}.
    
    Model calculations predict that the yields of \hl and \hel should be similar in heavy-ion collisions~\cite{Steinheimer:2012tb,Glassel:2021rod}. However, the number of analyzed \hel is much less than the number of analyzed \hl due to the lower acceptance in STAR for three-body decays, leading to the statistical uncertainty on the \hel mass driving the statistical uncertainties on the $\Lambda$ binding-energy differences. Besides, the $\Lambda$ binding energy difference between \hl and \hel from the experiments in the 1970s was measured both in their three-body decay channels~\cite{NPB1973}. To compare with it, it may be more reasonable for us to address the CSB effect also in their three-body decay channels, which requires a reconstruction of \hl via its three-body decay channel $\rm ^4_{\Lambda}H~$$\to t+p+\pi^-$. However, the three-body decays have lower acceptance than two-body decays in STAR and a smaller branching ratio~\cite{STAR:2021orx}. Furthermore, due to the +1 charge of the triton, the $dE/dx$ of the triton usually mixes with other particles with +1 charge as shown in Fig.~\ref{fig1}. These conditions lead to the statistics of \hl reconstructed via the three-body decay channel being much lower than \hl two-body decay and \hel three-body decay. Therefore, we did not consider the three-body decay channel of \hl in this analysis. STAR has collected more statistics in the fixed-target mode. Within a few years for data production and analysis, the precision of current binding-energy measurements will be improved. The \hl three-body decay channel analysis may also become possible, and one may also have the chance to study the YNN interaction via the momentum correlation between $\Lambda$ and light nuclei~\cite{Haidenbauer:2021wld,Shao:2020lbq}. 
    
    \section{Summary}
    In summary, the masses and the $\Lambda$ binding energies of the mirror hypernuclei, \hl and \held, are measured in Au+Au collisions at $\sqrt{s_{\rm NN}}=3$~GeV. By using the $\gamma$-ray transition energies of the excited states from previous measurements~\cite{Bedjidian:1976zh,Yamamoto:2015avw}, the $\Lambda$ binding energies of them in excited states are also extracted. The CSB effect in $A = 4$ hypernuclei are then studied by measurements of the $\Lambda$ binding-energy differences between the ground states of \hl and \hel or their excited states. In comparison with other experimental measurements and theoretical studies, our results with a positive $\Delta B_{\Lambda}^4(0_{\rm g.s.}^{+})$ and a negative $\Delta B_{\Lambda}^4(1_{\rm exc}^{+})$ of comparable magnitudes within uncertainties, are consistent with the calculation using chiral effective field theory YN potentials plus a CSB effect. Although the statistical uncertainties are large, our approach provides a new avenue to study the CSB in heavy-ion collision experiments. 
    
    \section{Acknowledgement}
    We thank the RHIC Operations Group and RCF at BNL, the NERSC Center at LBNL, and the Open Science Grid consortium for providing resources and support.  This work was supported in part by the Office of Nuclear Physics within the U.S. DOE Office of Science, the U.S. National Science Foundation, National Natural Science Foundation of China, Chinese Academy of Science, the Ministry of Science and Technology of China and the Chinese Ministry of Education, the Higher Education Sprout Project by Ministry of Education at NCKU, the National Research Foundation of Korea, Czech Science Foundation and Ministry of Education, Youth and Sports of the Czech Republic, Hungarian National Research, Development and Innovation Office, New National Excellency Programme of the Hungarian Ministry of Human Capacities, Department of Atomic Energy and Department of Science and Technology of the Government of India, the National Science Centre of Poland, the Ministry of Science, Education and Sports of the Republic of Croatia, German Bundesministerium f\"ur Bildung, Wissenschaft, Forschung and Technologie (BMBF), Helmholtz Association, Ministry of Education, Culture, Sports, Science, and Technology (MEXT) and Japan Society for the Promotion of Science (JSPS).

\bibliography{myref}

\end{document}